\documentclass[twocolumn,         
               showpacs,longbibliography,            
               showkeys,preprintnumbers,     
               aps,                 
               prd,          	    
               letterpaper,             
               superscriptaddress,      
               nofootinbib,         
               tightenlines,        
               floats,floatfix      
               ]{revtex4-1}
\usepackage{physics}
\usepackage{epsf}
\usepackage{graphicx,color}
\usepackage{subfigure}
\usepackage{latexsym}
\usepackage{amsmath,amssymb}        
\usepackage[colorlinks=true,linkcolor=blue,citecolor=blue]{hyperref}
\usepackage{mathrsfs}
\usepackage{comment}
\usepackage{soul}
\usepackage{cancel}
\usepackage{graphicx}

\usepackage{bm}
\usepackage{tikz}

\usepackage{adjustbox}
\definecolor{purple}{rgb}{0.58,0.0,0.83}
\definecolor{orange}{rgb}{1,0.5,0}
\DeclareSymbolFontAlphabet{\mathrsfs}{rsfs}
\DeclareMathAlphabet{\mathcal}{OMS}{cmsy}{m}{n}

\newcommand\bracketPaco[2]{\langle #1 | #2 \rangle} 
\newcommand\sandwichPaco[3]{\langle #1 |#2| #3 \rangle}

\newcommand{\circledplus}{\mathbin{\tikz[baseline=-0.6ex]{
            \node[draw, circle, inner sep=0.5pt] (o) {\scalebox{0.7}{$+$}};
        }}}
        
\newcommand{\circledminus}{\mathbin{\tikz[baseline=-0.6ex]{
            \node[draw, circle, inner sep=0.5pt] (o) {\scalebox{0.7}{$-$}};
        }}}

\begin{document}


\title{Variational Quantum Crank-Nicolson and Method of Lines for the Solution of Initial Value Problems}


\author{Francisco Guzman-Cajica}
\email{2002236d@umich.mx}
\affiliation{Facultad de Ciencias F\'isico Matem\'aticas, Universidad
              Michoacana de San Nicol\'as de Hidalgo. Edificio ALFA, Cd.
              Universitaria, 58040 Morelia, Michoac\'{a}n,
              M\'{e}xico.}  

\author{Francisco S. Guzm\'an}
\email{francisco.s.guzman@umich.mx}
\affiliation{Instituto de F\'{\i}sica y Matem\'{a}ticas, Universidad
              Michoacana de San Nicol\'as de Hidalgo. Edificio C-3, Cd.
              Universitaria, 58040 Morelia, Michoac\'{a}n,
              M\'{e}xico.}  


\date{\today}


\begin{abstract}
In this paper we use a Variational Quantum Algorithm to solve Initial Value Problems with the Implicit Crank-Nicolson and the Method of Lines (MoL) evolution schemes. The unknown functions use a spectral decomposition with the Fourier basis. The examples developed to illustrate the implementation are the Advection equation, the wave equation written as a system of first order coupled equations and the viscous Burgers equation as a non-linear case. The problems are solved using: i) standard Finite Differences as the solution to compare with, ii) the State Vector Formalism (SVF), and iii) the Sampling Error Formalism (SEF). The contributions of this paper include: 1) cost functions for generic first order in time PDEs using the implicit Crank-Nicholson and the MoL, 2) detailed convergence or self-convergence tests are presented for all the equations solved, 3) a system of three coupled PDEs is solved, 4) solutions using sampling error are presented, which highlights the importance of simulating the sampling process and 5) a fast version of the SVF and SEF was developed which can be used to test different optimizers faster.
\end{abstract}




\maketitle

\section{Introduction}
\label{sec:intro}

Among the most important applications of Computer Science is the solution of Partial Differential Equations (PDEs) that serve to model a garden variety of dynamical processes in nature. This application is expected to be continued in the era of quantum computers under development; the implementation of numerical quantum algorithms for this purpose is therefore an active area of research. 
Variational Quantum Algorithms (VQA) are a class of hybrid algorithms that aim to solve optimization problems by minimizing cost functions that depend on a set of parameters. The cost functions are calculated using a quantum computer and the space of parameters is explored using a classical one \cite{NatureVQA}. The idea of applying VQAs to solve PDEs is under current development for a number of well known equations. For example, in \cite{Lubasch_2020} a VQA is used to solve the stationary non-linear Schr\"odinger equation, in \cite{AppPlasmaPhysics} the methods are applied to certain plasma physics problems, in  \cite{BlackScholes} the Black-Scholes equation for the calculation of financial risk is solved, in \cite{AppHeatEquation} applications to heat transfer are implemented, in  \cite{AppPoisson} Poisson type equations are solved, in \cite{AppNavierStokes} Navier-Stokes equations are solved for a 2D fluid, in \cite{AppColloidal} Colloidal Transport is simulated through the Smoluchowski equation, and even in astrophysics the first steps in the application to simulations of dark matter are already on course \cite{Mocz_2021,AppSP2024}. This set involves both Initial Value Problems (IVP) as well as Boundary Value Problems, which allows one to foresee an interesting potential applicability.

As far as we can tell, there is not a consolidated combination of ingredients of the VQA, which include: the ansatz assumed for the unknown function, the evolution scheme, and the optimization algorithm. Instead, it seems that these elements depend on the type of PDE to be solved. This is the reason why different evolution, parametrization and optimization methods are being explored wildly as in the examples mentioned above.

The aim of this paper is to implement a VQA in the solution of Initial Value Problems in 1+1 dimensions based on two of the most traditional Finite Differences evolution schemes, specifically the implicit Crank-Nicolson and the Method of Lines (MoL), which are widely used in the classical numerical solution of IVPs. We illustrate the functioning of these methods with three traditional problems: i) the Advection equation, which is a very simple introductory case, ii) the wave equation used to show how to implement the solution of systems of various coupled equations, and iii) the viscous Burgers equation to illustrate the solution of non-linear IVPs that tend to develop sharp solutions. 

Since we are interested in the implementation of the evolution schemes and the convergence of solutions, we solve the examples in a periodic domain and avoid the complications due to boundary conditions. Thus our implementation uses a spectral decomposition of the unknown function using the Fourier basis following the recipe in \cite{princeton,BlackScholes}. The accuracy of the VQA is studied by simulating the quantum computation of the cost functions in a classical computer. This is done in two ways: 1) using the State Vector Formalism (SVF) that computes the complete wave function of the quantum circuits to get an exact probability distribution, and 2) the Sampling Error Formalism (SEF), which finds approximated probability distributions simulating Monte Carlo errors, and provides an idea of how these methods would perform in an ideal quantum computer.

This paper contributes with the cost functions required to implement the implicit Crank Nicholson and the Method of Lines (MoL) for a generic first order in time equation. We also present detailed convergence tests of the numerical solutions, which help evaluate consistency and convergence properties of the solutions using different numbers of qubits. Specifically, we practice time-dependent \textit{convergence} tests for the cases with exact solution, and \textit{self-convergence} tests for the case without exact solution. These tests help confirm or discard the validity of the various numerical solutions, and also help determine the convergence regime in terms of the number of qubits. Other contributions of this paper include a fast version of the SVF and SEF which work specifically for the spectral decomposition used, but that can be advantageous to test out different optimizers. The wave equation is a particular example that we use to describe the MoL for a system of coupled first order equations. Since this is an explicit method, the cost function can be straightforwardly generalized to a system of an arbitrary number of equations. Finally, a complexity analysis of the algorithm is conducted to show the improvement against the classical numerical methods.

Our results indicate that the spectral decomposition produces solutions that converge with second order in a given time-window for the Advection and wave equations, when using more than 3 qubits. In the case of the viscous Burgers equation we found a self-convergence between first and second order in a finite time-window also with more than 3 qubits. We implemented the SEF, which is usually avoided in quantum simulations and shows a closer idea of how a quantum computer would perform. Despite the stochastic cost functions  which result from the circuit sampling, the solutions found are similar and can be considered and approximation of the SVF calculation.

The paper is organized as follows. 
In Section \ref{sec:method} we describe the two classical Finite Differences methods to be exemplified.
In Section \ref{sec:quantummethod} we develop the generic quantum version of each of them.
In Section \ref{subsec:advection} we present the solution of the Advection equation that illustrates how to handle a Crank-Nicolson time-average, 
in Section \ref{subsec:wequation} we show the implementation of the MoL for the solution of the wave equation, and 
in Section \ref{subsec:bequation} we solve the viscous Burgers equation. 
A complexity analysis of our approach is found in Section \ref{sec:analysis}. Finally, in Section \ref{sec:conclusions} we draw some conclusions and final comments.

\section{Traditional Methods}
\label{sec:method}

In this work, we solve evolution equations in a 1+1 domain $D$ characterized by one spatial dimension described by the coordinate $x$ and time $t$. Explicitly $D:=[x_{\rm min},x_{\rm max}]\times[0,t_{\rm f}]$. Since we are focused on the implementation and testing of evolution schemes, in all the examples below a periodic domain is used. In general, an evolution equation can often be written as: 

\begin{equation}
    \partial_t u = {\rm rhs}(u,\partial_x u, \partial_{xx} u, ...),
\label{eq:differentialEquation}
\end{equation}

\noindent for some initial conditions $u(x,0)=u_0(x)$. The right hand side (rhs) of this equation can be any function of $t$, the unknown $u$ and  its derivatives. The first step to calculate the approximate solution is to define the discrete domain $D_d$ as the set of points $\{(x_i,t^n)\}$, where $x_i = x_{\rm min}+i\Delta x$ for $i=0,1,\dots,N-1$, $\Delta x = (x_{\rm max}-x_{\rm min}) /N$ is the spatial resolution, $t^n=n\Delta t$ and $\Delta t = {\rm CFL}~ \Delta x$. Here the constant ${\rm CFL}$ is the Courant Friedrichs Lewy factor, a parameter that often restricts the stability of an evolution scheme.

The unknown continuous function $u(x,t)$ is reduced to a set of real numbers $u_i^n$ which correspond to the value of $u(x_i,t^n)$. These values are stored in a vector $\vec{u}^n = \vec{u}(t^n) = (u_0^n,\dots,u_{N-1}^n)$, and the continuous version of the differential equation is reduced to local algebraic equations. The way the equation is discretized depends on the particular method being used. However,  first and second derivatives are always replaced by discrete expressions like:

\begin{eqnarray}
\partial_x u (x_i)&=& \frac{1}{2\Delta x}[u_{i+1} - u_{i-1}] + {\cal O}(\Delta x^2),
\nonumber\\
\partial_{xx} u (x_i) &=& \frac{1}{\Delta x^2}[u_{i+1} -2u_i + u_{i-1}] + {\cal O}(\Delta x^2). 
\nonumber
\end{eqnarray}

\noindent which are second order accurate. It is sometimes useful to describe these operators with vectors like $\partial_x\vec{u} = (\partial_x u (x_0),\dots,\partial_x u (x_{N-1}))$, which can be described with the following notation:

\begin{eqnarray}
\partial_x \vec{u} &\simeq& \frac{1}{2\Delta x} (\circledplus - \circledminus )~\vec{u}, \nonumber\\ 
\partial_{xx} \vec{u} &\simeq& \frac{1}{\Delta x^2} (\circledplus -2\mathbf{I} + \circledminus)\vec{u},\nonumber
\end{eqnarray}

\noindent where the linear unitary operators $\circledplus$ and $\circledminus$ are $i-$label periodic shifters and will show helpful later on, within the quantum algorithm. In what follows, two evolution methods based on finite differences will be described.

\subsection{Implicit Crank-Nicolson}
\label{subsec:Dd}

In this method, Eq. (\ref{eq:differentialEquation}) is replaced by the Crank-Nicolson time average approximation given by

\begin{equation}
\frac{u_i^{n+1} - u_i^{n}}{\Delta t} = \frac{1}{2} \left[ {\rm rhs}_i^{n+1} + {\rm rhs}_i^n \right],
\label{eq:CNgenericOriginal}
\end{equation}

\noindent where ${\rm rhs}_i^{n+1}$ and ${\rm rhs}_i^{n}$ are, in our case, second order accurate discretized versions of the right hand side of the general equation (\ref{eq:differentialEquation}) at the spatial point $x_i$, that may contain derivatives of the unknown. The whole formula is second order accurate at the point $(x_i,t^{n+1/2})$, which does not belong to $D_d$ but serves to calculate $u^{n+1}_i$. This last step is carried out by solving a linear system of equations in traditional calculations \cite{nr}. For our purpose, we write Eq. (\ref{eq:CNgenericOriginal}) for all $i$ with vector notation:

\begin{equation}
\frac{\vec{u}^{n+1} - \vec{u}^{n}}{\Delta t} = \frac{1}{2} \left[ \vec{\rm rhs}^{n+1} + \vec{\rm rhs}^n \right],
\label{eq:CNgeneric}
\end{equation}

\noindent which will ease the conversion to quantum notation.

\subsection{Method of Lines}
\label{sec:QMoL}

Unlike the Crank-Nicolson, the MoL is a generalized version of explicit methods that focus on the solution along the time direction at the spatial location $x_i$. Given an IVP with the evolution equation (\ref{eq:differentialEquation}) 
and initial data $u_0(x)$, the idea is to formulate a semi-discrete version of this problem at point $(x_i,t^n)\in D_d$ as follows:

\begin{eqnarray}
\partial_t u |_{(x_i,t^n)} &=& {\rm rhs}(u^n_{i},{\rm DV}(\partial_x u)|_{(x_i,t^n)},\nonumber\\
&&{\rm DV}(\partial_{xx} u)|_{(x_i,t^n)},\dots),\nonumber
\end{eqnarray}

\noindent where ${\rm DV}$ stands for the Discrete Version of derivative operators of desired accuracy. The point is that, given a discrete version of ${\rm rhs}$, this expression defines a set of $N$ coupled ODEs that can be integrated using the desired ODE integrator. The accuracy and stability of the method depend on the accuracy of the spatial operators that approximate spatial derivatives  as well as the time integrator used. Specifically, in this paper we use second order accurate stencils to approximate spatial derivatives and for time-integration the Heun flavor of the second order accurate Runge-Kutta that we now describe. 

The explicit two step implementation of this method at point $x_i$ to evolve $u$ from time $t^n$ to $t^{n+1}$ reads (see e. g. \cite{fsguzman}):

\begin{eqnarray}
u^*_i &=& u^n_i + \Delta t ~{\rm rhs}(u^n_i),\label{eq:RK2a}\\
u^{n+1}_i &=& \frac{1}{2}\left[u^n_i + u^*_i + \Delta t ~{\rm rhs}(u^*_i) \right],\label{eq:RK2b}
\end{eqnarray}

\noindent which is the consecutive application of two explicit Euler steps that we label $\text{RK2-}a$ and $\text{RK2-}b$ respectively.

\section{Quantum methods}
\label{sec:quantummethod}

The VQA has three main parts: i) the anzats, which defines the representation of the unknown function $u$ in terms of a set of parameters, for which we use the Fourier basis as described in \cite{princeton}; ii) the definition and calculation of a cost function as described in \cite{Lubasch_2020}; iii) the optimization of the parameters done in a classical computer with standard optimizers from {\tt scipy}, in our case the {\tt Nelder-Mead}.

In the discrete classical method, the function $u(x)$ is replaced by a vector of real entries $\vec{u}$. The first step behind the VQA is to codify this vector in a set of ${\rm n}$ qubits, where ${\rm n} = \log_2(N)$. The state of a system of ${\rm n}$ qubits, is represented by a wave function. This wave function is a linear combination of vectors from a basis. Following \cite{libroDeKinder}, the canonical basis is composed of vectors of the form: $\ket{abc\cdots}:=\ket{a}\otimes\ket{b}\otimes\ket{c}\otimes\cdots := \ket{2^{{\rm n}-1}a+2^{{\rm n}-2}b+2^{{\rm n}-3}c+\cdots }$, where $a,b,c,\cdots$ take values 0 or 1. An example that helps understanding the circuits below is the following. Consider the state $\ket{100}$ which is the tensor product of the individual states of each qubit: $\ket{1}\otimes\ket{0}\otimes\ket{0}$. This state means that the first qubit is in the state $\ket{1}$, the second in the state $\ket{0}$ and the third in the state $\ket{0}$. To simplify the notation, we would call such a state $\ket{4}$, because $100$ is the binary representation of the number $4$. With this in mind, a general state of the set of ${\rm n}$ qubits can be written as:

\begin{equation}
    \ket{\psi} = \sum_{i=0}^{N-1} \psi_i \ket{i}, \label{eq:psiexpansion}
\end{equation}

\noindent where the $\psi_i$ are complex numbers called probability amplitudes. Their physical meaning is that, when measured, the system has a probability $|\psi_i|^2$ of collapsing to the state $\ket{i}$. One interesting fact about this system is that the dimension of the Hilbert space where $\ket{\psi}$ lives, grows exponentially with the number of qubits ${\rm n}$. One first idea would be to encode the $u_i$ components of the original vector into a quantum state as follows:

\begin{equation}
    \ket{u(t^n)} = \sum_{i=0}^{N-1}u_i^n \ket{i}. \nonumber
\end{equation}

\noindent One problem with this ket is that it is not normalized. For this reason, it cannot really be the wave function of a system of ${\rm n}$ qubits. To fix this, we can simply pull out a factor and define: 

\begin{equation}
    \ket{\psi} = \sum_{i=0}^{N-1} \psi_i \ket{i} = \frac{1}{\lambda_0}\sum_{i=0}^{N-1}u_i^n \ket{i}.\nonumber
\end{equation}

\noindent This state can now effectively be encoded in the wave function of an ${\rm n}$ qubit system. 

\subsection{The ansatz}

Setting up such a quantum state in a set of ${\rm n}$ qubits is possible, but requires an exponentially growing number of gates as ${\rm n}$ grows. For that reason, we will only approximate the $\psi_i$ components. We will define a unitary operator $\hat{U}$ which depends on a vector of parameters $\bm{\lambda}$. This operator will allow us to create sufficiently general wave functions with a number of parameters that grows logarithmically with the number of points of the discrete domain $D_d$. That is, if the discrete domain has $N$ points, the idea is to only require $O({\rm poly}(\log{N})) = O({\rm poly}({\rm n}))$ of parameters. We call this $\hat{U}$ operator, the ansatz. This operator has the following property:

\begin{equation}
    \ket{u(\bm{\lambda})} = \lambda_0\hat{U}(\bm{\lambda})\ket{0},\label{eq:expansionpsi}
\end{equation}

\noindent that is, it turns the ground state of the system into a state whose wave function approximates the normalized $\vec{u}$ vector we want to encode. Also see that $\lambda_0$ not only works as a normalization constant, capable of turning the components of a normalized vector into the $u_i^n$ values, but also as a free parameter. With this in mind, the approximate solution will be given by the state $\ket{u(\bm{\lambda}(t))}$. In other words, instead of finding the $u$ function at every time step as in the classical calculation, we only need to find the $\bm{\lambda}$ parameters at the discrete times $t^n$ that produce the $\ket{u(t^n)}$ that best satisfies the discrete version of the evolution equation.

There are different options for the $\hat{U}$ operator. For the implementations in this work, we use a truncated Fourier series expansion as the ansatz. This idea is bassed in the Fourier Series Loader introduced in \cite{princeton,BlackScholes}. The motivation behind this choice is that, since we are interested in the evolution schemes, we will only treat periodic boundary conditions for the IVPs. Therefore, the $u(x)$ function will be periodic and can be expressed as a Fourier series. Additionally, this ansatz has the advantage that, when the parameters are varied, the $u(x)$ function changes smoothly. This behaviour is absent in other ansatz such as the ones used in \cite{AppColloidal,AppNavierStokes,Mocz_2021,Lubasch_2020}.

We start by representing the $\psi_i$ components from Eq. (\ref{eq:psiexpansion}) in terms of the Fourier coefficients:
\begin{equation}
    \psi_i = \sum_{p = -M}^M c_p e^{-{\rm i}2\pi p i/N},
    \label{eq:spectral}
\end{equation}

\noindent where $M$ is the number of Fourier frequencies to be included in the series. To codify this series in the quantum computer, ${\rm m}+1$ qubits are used to store the amplitudes of each Fourier mode. The condition that must be satisfied by m is $2^{{\rm m}+1} \ge 2M+1$. This implies that $M \le 2^{\rm m} - 1/2$ but, since $M$ has to be an integer, $M \le 2^{\rm m} - 1$. Therefore, from the n qubits used to codify the $|\psi\rangle$ state, a subset of ${\rm m}+1$ qubits are employed to momentarily store the amplitudes of the $2M+1$ Fourier modes. In this paper we choose $M = 2^{\rm m} -1$ following the steps in \cite{princeton}, but $M$ can in principle be smaller.
On the other hand, the $\bm{\lambda}$ vector is the array of the $2M+1$ coefficients $c_p$. The process of setting up the quantum state in terms of the $c_p$ coefficients is done efficiently in \cite{princeton}, and the authors share a python library with the implementation that we do not need, but that could be useful when implementing the VQA on a real quantum computer \cite{princetonlib}. 

Now that $\bm{\lambda}$ was been defined, the only key idea that we need to know about the internal structure of the $\hat{U}$ gate is that it has the form:

\begin{equation}
    \hat{U}(\bm{\lambda}) = {\rm QFT}^{\dagger}~ \hat{V}(\bm{\lambda}),\label{eq:UqftV}
\end{equation}

\noindent that is, the ansatz consists of a product of two gates, $\hat{V}(\bm{\lambda})$ that depends on the parameters, followed by an inverse Quantum Fourier Transform. This property will help simplify upcoming quantum circuits.

This ansatz will be used for both, setting up the initial condition of the function and to codify the function at later times. To find the $\bm{\lambda}$ parameters at initial time time $t^0$, we use the following procedure:

\begin{itemize}
    \item[-] Chose the initial condition $u_0(x)$.
    \item[-] Evaluate the function in the discrete domain to get the vector $\vec{u} = (u_0, u_1,\dots,u_{N-1})$.
    \item[-] Normalize the $\vec{u}$ vector to obtain $(1/\lambda_0)\vec{u}$, saving the normalization factor $\lambda_0$.
    \item[-] Find the Fast Fourier Transform of $(1/\lambda_0)\vec{u}$. The first $M+1$  and the last $M$ entries of the transform correspond to the coefficients of the modes with lowest frequency. These entries will be the $c_p$ coefficients and constitute the vector of parameters $\bm{\lambda}(t^0)$.
\end{itemize}

\noindent With this $\bm{\lambda}(t^0)$ vector, we can now set the initial condition $u_0(x)$ from (\ref{eq:differentialEquation}) in the quantum circuits.

\subsection{The Cost Function}
\label{subsubsec:costfunction}

Once $\bm{\lambda}(t^n)$ is known, the objective is to find the parameters $\bm{\lambda}(t^{n+1})$ that best solve the evolution equation at time $t^{n+1}$. In the rest of the section, we will focus in finding the $\bm{\lambda}(t^{n+1})$ that we denote by $\lambda$, in terms of $\bm{\lambda}(t^{n})$ denoted by $\tilde{\lambda}$. We also define $\ket{u} := \ket{u(\lambda)}$ and $\ket{\tilde{u}} := |u(\tilde{\lambda}) \rangle$ as the states at times $t^{n+1}$ and $t^{n}$ respectively. With these conventions, the second step in the VQA is to define a {\it Cost Function.} We need a way to know how the $\bm{\lambda}$ parameters change. This can be done using any of the finite difference methods described above. 

We start by defining the cost function for the implicit {\it Crank-Nicolson}. This evolution scheme has already been applied to the solution of the diffusion-reaction and Navier-Stokes equations, in \cite{AppNavierStokes}, although in what follows we develop a considerably different cost function that should work for a general first order in time evolution equation. According to Eq. (\ref{eq:CNgeneric}), the following state:

\begin{equation}
|EQ\rangle:=|u\rangle - |\tilde{u}\rangle - \frac{\Delta t}{2}[ | {\rm rhs}\rangle + |\tilde{\rm rhs}\rangle], \nonumber
\end{equation}

\noindent measures the violation of the discrete version of the equation, and should be zero if the equation is exactly fulfilled. However this is not the case in any numerical calculation, instead one looks for the state $|u(\lambda)\rangle$ that makes $\ket{EQ}$ the smallest. For this reason, the cost function will be defined as the squared norm of such ket:

\begin{eqnarray}
CF_{CN} &:=& \langle EQ | EQ \rangle = \nonumber\\
&&\left( |u\rangle - |\tilde{u}\rangle - \frac{\Delta t}{2}[ | {\rm rhs}\rangle + |\tilde{\rm rhs}\rangle] \right)^{\dagger} \big| \nonumber\\
&&\left( |u\rangle - |\tilde{u}\rangle - \frac{\Delta t}{2}[ | {\rm rhs}\rangle + |\tilde{\rm rhs}\rangle] \right) \nonumber\\
&=& \braket{u}{u} - 2 {\rm Re}\{ \bracketPaco{\tilde{u}}{u} \}\nonumber\\
&-& \Delta t {\rm Re}\{\bracketPaco{\rm rhs}{u}+\bracketPaco{\tilde{\rm rhs}}{u}-\bracketPaco{\tilde{u}}{\rm rhs}\} \nonumber\\
&+& \frac{\Delta t^2}{4}\bracketPaco{\rm rhs}{\rm rhs} + \frac{\Delta t^2}{2} Re\{\bracketPaco{\tilde{\rm rhs}}{\rm rhs}\} \nonumber\\
&+& {\rm constant},\label{eq:CFgeneralCN}
\end{eqnarray}

\begin{figure}
\includegraphics[width=8.5cm]{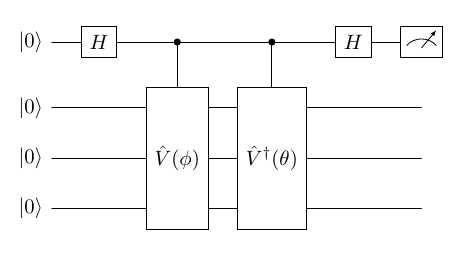}
\caption{Circuit that evaluates the product ${\rm Re} \{ \langle 0 | \hat{U}^{\dagger}(\theta) \hat{U}(\phi) |0\rangle\}$ typically found in our cost functions. In this example ${\rm n} = 3$. The circuit diagrams in this paper were constructed using the {\tt quantikz} package introduced in \cite{quantikzTutorial}.}
\label{circuit:dotProduct}
\end{figure}

\begin{figure}
\includegraphics[width=8.5cm]{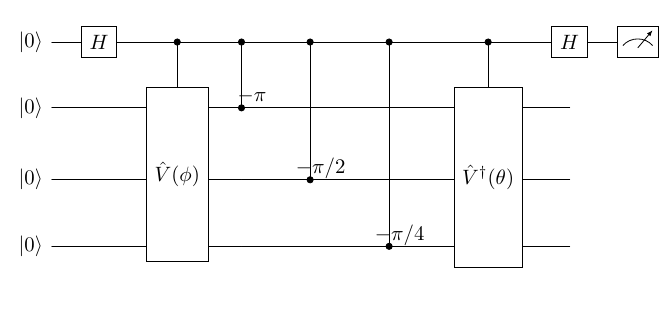}
\caption{Circuit that evaluates the term $ {\rm Re}\{ \langle 0 | \hat{U}^{\dagger}(\theta)\oplus \hat{U}(\phi) | 0 \rangle $, which typically appears in cost functions. In this example ${\rm n} = 3$.}
\label{circuit:dotWithPlus}
\end{figure}

\noindent where the constant at the end is a function which only depends on $\tilde{\lambda}$ and does not depend on $\lambda$. The idea of the VQA is to compute this cost function efficiently with a quantum computer.

The complete cost function is complicated, it has a lot of dot products and expectation values. Usually individual products are computed and are then added up. As an example, one common product that appears in our cost functions is of the type ${\rm Re} \{ \bracketPaco{u(\theta)}{u(\phi)}\}$, where $\theta$ and $\phi$ are two vectors of parameters. If we use equations (\ref{eq:expansionpsi}) and (\ref{eq:UqftV}), this product becomes:
{\small
\begin{eqnarray}
    {\rm Re} \{ \bracketPaco{u(\theta)}{u(\phi)}\} &=& {\rm Re} \{\theta_0^* \phi_0 \langle 0 | \hat{U}^{\dagger}(\theta) \hat{U}(\phi) |0\rangle\} \nonumber \\
    &=& {\rm Re} \{\theta_0^* \phi_0 \langle 0 | \hat{V}^{\dagger}(\theta)~{\rm QFT}~{\rm QFT}^{\dagger}~\hat{V}(\phi) |0\rangle\} \nonumber \\
    &=& {\rm Re} \{\theta_0^* \phi_0 \langle 0 | \hat{V}^{\dagger}(\theta)\hat{V}(\phi) |0\rangle\}, \nonumber
\end{eqnarray}}

\noindent which is nothing more than the real part of the expectation value of the operator $\hat{V}^{\dagger}(\theta)\hat{V}(\phi)$ in the state $\ket{0}$, scaled by the factor ${\rm Re} \{\theta_0^* \phi_0\} = \theta_0 \phi_0$. To calculate the real part of the expectation value of a unitary operator, a Hadamard Test can be employed. The circuit of such test for this particular product can be seen in Figure \ref{circuit:dotProduct}. 

The circuit consists of one auxiliary control qubit, called ancilla qubit, and n qubits for the storage of the wave function. Every time the circuit is ran, the ancilla qubit has a $P_0$ probability of collapsing to the state $\ket{0}$ and a $P_1$ probability of collapsing to the state $\ket{1}$. The grace of this test is that $P_0 - P_1 := \langle \hat{\sigma}_z\rangle$ happens to be equal to ${\rm Re} \{ \langle 0 | \hat{V}^{\dagger}(\theta)\hat{V}(\phi) |0\rangle\}$. In this case $\hat{\sigma}_z$ is the $z-$Pauli matrix. So, in order to calculate the product, this circuit has to be ran multiple times: $T$ times. Collecting the results of the ancilla qubit, one can retrieve approximated values $P_{0_{\rm approx}} = \text{\# zeros}/T$ and $P_{1_{\rm approx}} = \text{\# ones}/T$ and calculate $P_{0_{\rm approx}}-P_{1_{\rm approx}} := \langle \hat{\sigma}_z\rangle_{\rm approx}$. However, this approximation differs from the exact value $\langle \hat{\sigma}_z\rangle$ by a Monte Carlo error that goes as \cite{Lubasch_2020}:
\begin{equation}
    |\langle \hat{\sigma}_z\rangle - \langle \hat{\sigma}_z\rangle_{\rm approx}| \simeq \frac{\sqrt{1-\langle \hat{\sigma}_z\rangle^2}}{\sqrt{T}}, \nonumber
\end{equation}

\noindent which decreases as $1/\sqrt{T}$. Here we call {\it sampling} to the process of repeating the measurement multiple times, and {\it sampling error} to the Monte Carlo error.

The other type of products that appear frequently in cost functions are of the form ${\rm Re} \{ \langle u(\theta) |\circledplus|u(\phi) \rangle \}$. The $\circledplus$ operator is a unitary gate that shifts the wave function's spatial label.  When acting on the basis, the result is $\circledplus \ket{i} = \ket{i-1}$. It can be shown that:

\begin{equation}
    {\rm QFT}\circledplus{\rm QFT}^{\dagger} = {\rm Diag}(1,e^{-{\rm i}2\pi/N},\dots,e^{-{\rm i}(N-1)2\pi/N}), \nonumber
\end{equation}

\noindent which is a diagonal matrix. With this result, the product simplifies to:

\begin{eqnarray}
    &&{\rm Re} \{ \langle u(\theta) |\circledplus|u(\phi) \rangle \} = \theta_0 \phi_0 {\rm Re} \{ \langle 0 | \hat{V}^{\dagger}(\theta) \nonumber\\
    &&~~~~~~~~ {\rm Diag}(1,\dots,e^{-{\rm i}(N-1)2\pi/N}) \hat{V}(\phi) |0\rangle\}. \nonumber
\end{eqnarray}

\noindent This product, without the $\theta_0 \phi_0$ factor, can be computed with the circuit of Figure \ref{circuit:dotWithPlus} measuring the ancilla qubit multiple times. Notice that the diagonal operator is composed of single qubit phase sifters. In general, if the system has ${\rm n}$ qubits, the phases must be: $-\pi,-\pi/2,\dots,-\pi/2^{({\rm n}-1)}$. One powerful feature of this circuit is that, if the phases are doubled, the resulting product is ${\rm Re} \{ \langle 0 | \hat{U}^{\dagger}(\theta) \circledplus^2 \hat{U}(\phi) |0\rangle\}$, and if the phases are multiplied by $-1$, the resulting product is ${\rm Re} \{ \langle 0 | \hat{U}^{\dagger}(\theta) \circledminus \hat{U}(\phi) |0\rangle\}$. Therefore, all these products can be computed using the same circuit with slightly modified phases.

Let us now refer to the {\it Method of Lines} implemented specifically with the Heun integrator defined in Eqs. (\ref{eq:RK2a})-(\ref{eq:RK2b}). In order to adapt the VQA, we define three states: $|\tilde{u} \rangle = |u(\tilde{\lambda}) \rangle$ at time $t^n$, $|u^* \rangle = |u(\lambda^*) \rangle$ an auxiliary intermediate state, and $|u \rangle = |u(\lambda) \rangle$ the state at time $t^{n+1}$. We first find the $\lambda^*$ parameters by minimizing the cost function corresponding to Eq. (\ref{eq:RK2a}):
\begin{equation}
    CF_{\text{RK2-}a} = \bracketPaco{u^*}{u^*} - 2 {\rm Re} \{ \bracketPaco{u^*}{\tilde{u}} \} - 2\Delta t {\rm Re} \{ \bracketPaco{u^*}{\tilde{{\rm rhs}}}\},
    \label{eq:CFgeneralRK2a}
\end{equation}

\noindent where $|\tilde{{\rm rhs}} \rangle$ is the resulting state obtained after applying the {\rm rhs} operator to $|\tilde{u} \rangle$. After finding the state with this first Euler step, one can compute the state at time $t^{n+1}$ by finding the $\lambda$ parameters that best minimize the cost function corresponding to Eq. (\ref{eq:RK2b}):
\begin{eqnarray}
    CF_{\text{RK2-}b} &=& \bracketPaco{u}{u} -  {\rm Re} \{ \bracketPaco{u}{\tilde{u}} + \bracketPaco{u}{u^*}\} \nonumber \\
    &-& \Delta t {\rm Re} \{ \bracketPaco{u}{{\rm rhs}^*}\},
    \label{eq:CFgeneralRK2b}
\end{eqnarray}

\noindent where $|{\rm rhs}^* \rangle$ is the state obtained from applying the rhs operator to $|u^* \rangle$. Most of the products that appear in the cost functions (\ref{eq:CFgeneralRK2a}) and (\ref{eq:CFgeneralRK2b}) can be computed using the circuits shown in Figures \ref{circuit:dotProduct} and \ref{circuit:dotWithPlus}.

\subsection{Minimization of Parameters}

The third and final step of the VQA is to minimize the cost function. We search for the $\bm{\lambda}$ parameters that minimize the various cost functions described above. After finding them, we would have completed one time step. This process is repeated at all time levels until final time.

There are various available optimizers for this purpose; however, some of them require the gradient or the Hessian matrix of the function that is being minimized. In this work, the computation of the derivatives of the cost function with quantum computers was not analyzed; therefore, a gradient free optimizer was needed. We chose the {\tt Nelder-mead} algorithm, which is capable of minimizing functions of many variables. If our function depends on $N_p = 2M + 1$ parameters, the algorithm uses a set of $N_p+1$ points in an $N_p$-dimensional space, to explore the space of parameters. This optimizer requires two evaluations of the function each time the parameters are modified.

\subsection{Implementation}

The methodology in this paper considers three ways to find the solution of the IVPs. The first one is the {\it Classical} evolution method, which solves the evolution using the plain and simple finite difference methods with a classical computer; this will be our numerical solution of reference, the best possible expected solution obtained numerically. The second one is the {\it State-Vector Formalism} (SVF), which uses the circuitry of the VQA by calculating the output wave function with a classical computer and retrieve the exact probability distribution of the ancilla qubit. The third one simulates that the ancilla qubit is being measured multiple times, which results in an approximate probability distributions more similar to what is expected in a quantum computer; we call this the {\it Sampling Error Formalism} (SEF).

There are various ways to simulate the quantum circuits in a classical computer, for example by using the {\tt State Vector} simulator and the {\tt Quasm} simulator provided by {\tt Qiskit} \cite{qiskit}. These simulators do not take into account quantum errors, like decoherence and decaying, which are present in today's quantum devices. However, we implemented the VQA with our own code that uses {\tt NumPy} array operations. The particular ansatz we are employing allows us to use the Fast Fourier Transform (FFT) to calculate the $\ket{u}$ state very efficiently. The function found in Eq. (\ref{eq:spectral}) is codified into a {\tt NumPy} array using the FFT instead of using matrix products as done by the {\tt State Vector} simulator. The dot products found in the cost functions can be computed using {\tt NumPy} dot products, which return the exact same output as the simulated circuits when using the {\tt State Vector} simulator from {\tt Qiskit}.

Monte Carlo sampling error can then be included in every dot product. For instance, if we want to simulate that we are computing ${\rm Re} \{ \sandwichPaco{0}{\hat{U}^{\dagger}(\theta)\hat{U}(\phi)}{0} \}$ by executing multiple times the quantum circuit and measuring the ancilla qubit, we would start by calculating $\hat{U}(\phi)|0\rangle$ and $\hat{U}(\theta)|0\rangle$ with the FFT. A {\tt NumPy} dot product can then be used to get $\langle \hat{\sigma}_z\rangle ={\rm Re} \{ \sandwichPaco{0}{\hat{U}^{\dagger}(\theta)\hat{U}(\phi)}{0} \}$, which is the expectation value of $\hat{\sigma}_z$ for the ancilla qubit. The $P_0$ and $P_1$ probabilities can be found solving the following system of equations:

\begin{eqnarray}
    P_0 - P_1 &=& \langle \hat{\sigma}_z\rangle, \nonumber\\
    P_0 + P_1 &=& 1, \nonumber\\
    \Rightarrow P_0 &=& \frac{1 + \langle \hat{\sigma}_z\rangle}{2}.\nonumber
\end{eqnarray}
Once $P_0$ is known, we can use the binomial distribution of {\tt NumPy} to simulate $T$ measurements of the ancilla qubit. With this information, the binomial distribution can provide the number of times that $0$ would appear if such $T$ experiments were to be performed. This provides the approximated probability distribution discussed in the previous subsection. This way of simulating the quantum circuits happens to be two orders of magnitude faster that the {\tt State vector} and {\tt Qasm} simulators provided by {\tt Qiskit} \cite{qiskit}. However, it must be kept in mind that this implementation can only be used with the Fourier ansatz we are using.

\section{Application of the Crank-Nicolson method: The Advection Equation}
\label{subsec:advection}

\begin{figure}
\centering
\includegraphics[width=8cm]{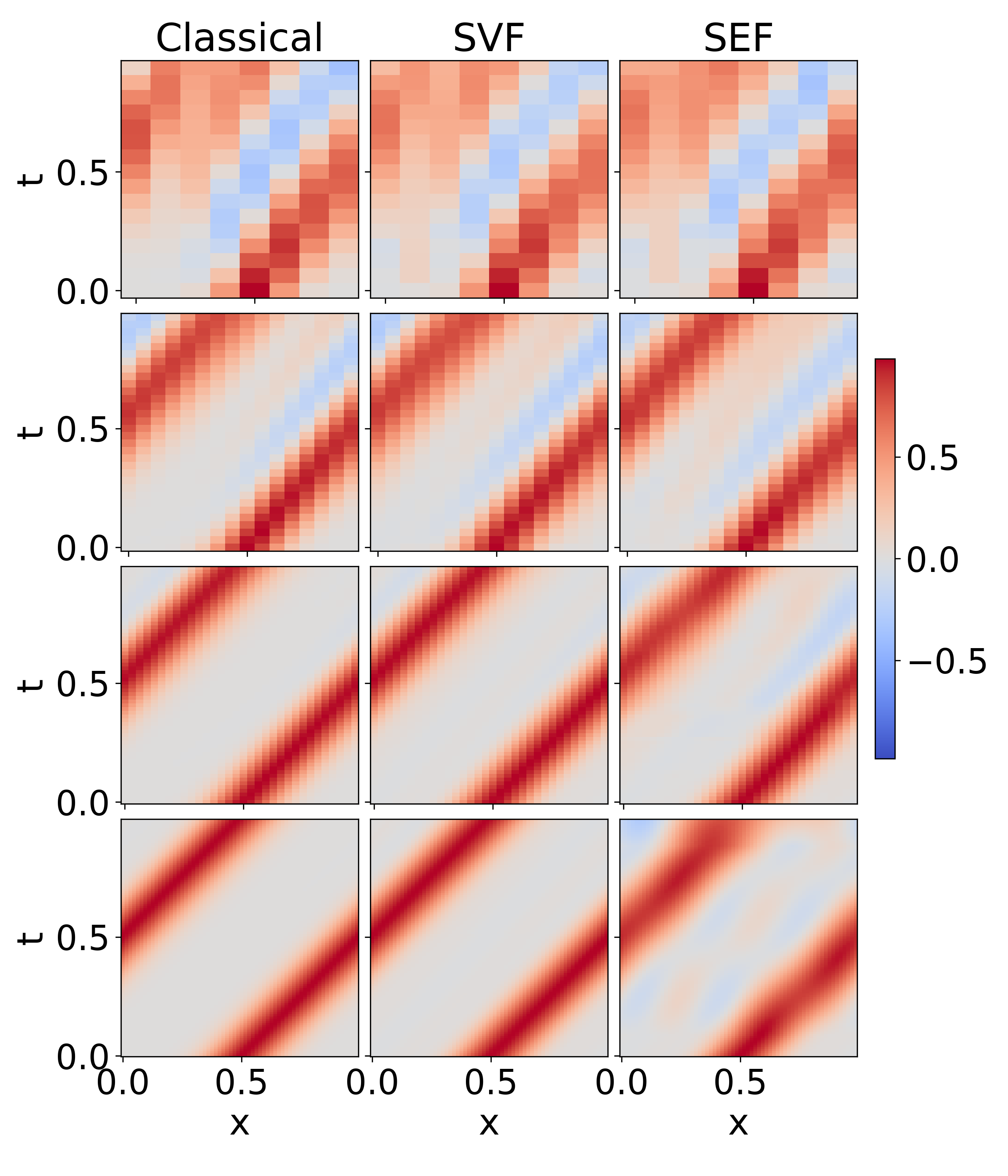}
\includegraphics[width=8cm]{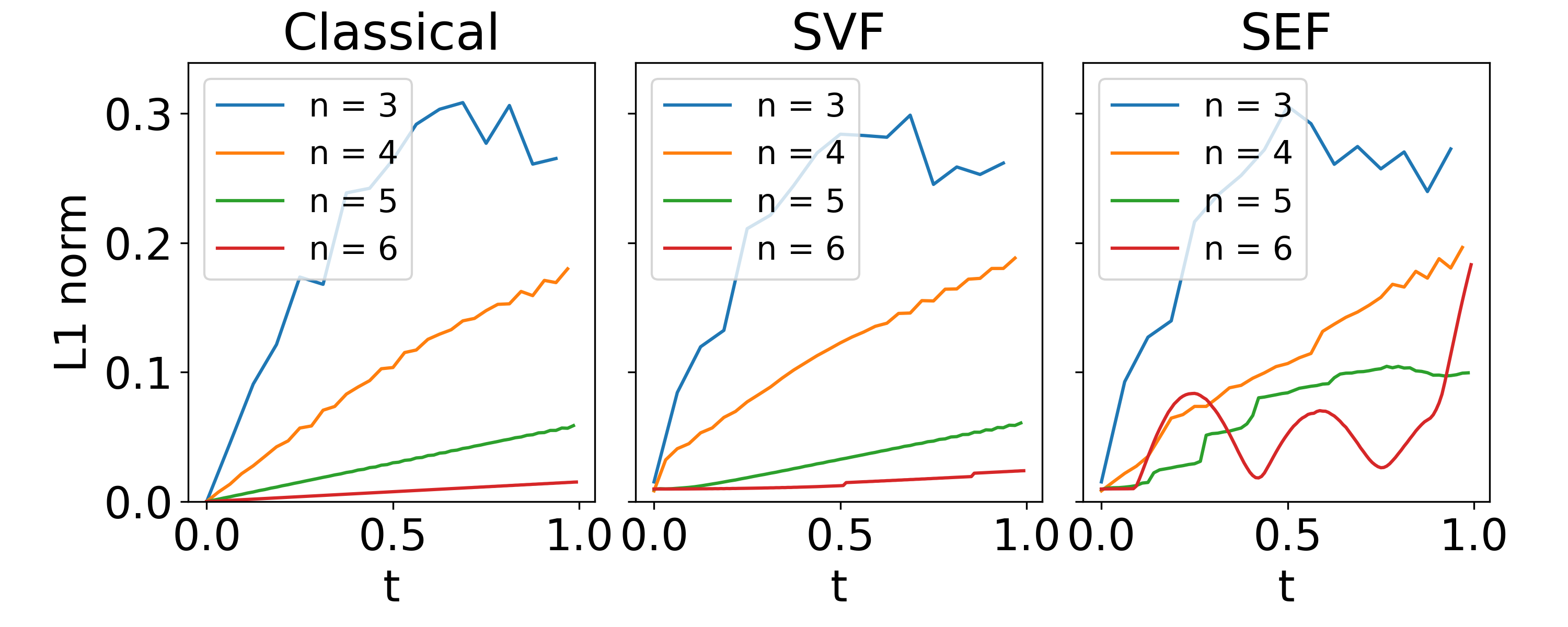}
\caption{From left to right we show the solution of the Advection equation using the Classical, SVF and SEF methods, respectively. At the first, second, third and fourth rows we show the solution calculated with 3, 4, 5 and 6 qubits respectively. The colors indicate the amplitude of the numerical solution $u_i^n$, an initially centered Gaussian of amplitude one that advects towards the right boundary and reenters from the left boundary. At the bottom we show a convergence test of the $L_1$ norm of the error as function of time using the four resolutions as described in the Appendix. This test indicates that the traditional method and the SVF converge with second order, whereas the SEF does not converge, instead it is very sensitive to evolution and strongly depends on the number of experiments; in this test we use $T=10^{11}$ experiments and still convergence is not achieved.}
\label{fig:ConvergenceAdvection}
\end{figure}

The first example to be solved is the Advection equation:

\begin{equation}
    \frac{\partial u}{\partial t} = - v \frac{\partial u}{\partial x}, \nonumber
\end{equation}

\noindent where $v$ is the propagation speed of initial data $u_0(x)$. The discrete version of the Crank-Nicolson time-average (\ref{eq:CNgeneric}) of this equation, using second order accurate derivative operators is:

\begin{equation}
    \frac{u^{n+1}_i - u^n_i}{\Delta t} = -v\frac{1}{2}\left[ \frac{u^{n+1}_{i+1} - u^{n+1}_{i-1}}{2\Delta x}+\frac{u^{n}_{i+1} - u^{n}_{i-1}}{2\Delta x} \right],\nonumber
\end{equation}

\noindent  which after simplifications leads to the following system of linear equations:

\begin{equation}
    \alpha u_{i+1}^{n+1} + u_{i}^{n+1} - \alpha u_{i-1}^{n+1} = -\alpha u_{i+1}^{n} + u_{i}^{n} + \alpha u_{i-1}^{n},\label{eq:TridiagonalAdvection}
\end{equation}

\noindent where $\alpha = 0.25v\Delta t/\Delta x$. In the traditional finite differences view, this defines a tridiagonal system of linear equations that we solve using traditional sparse matrix systems solvers \cite{nr}.

Concerning the VQA, the cost function (\ref{eq:CFgeneralCN}) corresponding to expression (\ref{eq:TridiagonalAdvection}) gives the following formula:

\begin{eqnarray}
CF &=& (1+2\alpha^2)\braket{u(\lambda)}{u(\lambda)}\nonumber\\
&+&(-2+4\alpha^2) {\rm Re} \{ \braket{u(\lambda)}{u(\tilde{\lambda})} \}\nonumber\\
&+&4\alpha {\rm Re}\{ \bra{u(\lambda)} \circledplus - \circledminus \ket{u(\tilde{\lambda})} \}\nonumber\\
&-& 2\alpha^2 {\rm Re} \{ \bra{u(\lambda)} \circledplus^2 \ket{u(\lambda)} \}\nonumber\\
&-& 2\alpha^2 {\rm Re} \{ \bra{u(\lambda)} \circledplus^2 + \circledminus^2 \ket{u(\tilde{\lambda})} \}
\label{eq:CFadvection}.
\end{eqnarray}

\noindent Each term of this expression is calculated using one of the two circuits described above and the results are then added up.

The solution of the Advection equation in the domain $D=[0,1]\times[0,1]$, for velocity $v=1$, with initial conditions $u_0(x)=e^{-(x-x_0)^2/\sigma^2}$, $x_0=0.5$ and $\sigma=0.15$, is summarized in Fig.  \ref{fig:ConvergenceAdvection}. We construct the solution for 3, 4, 5 and 6 qubits, which corresponds to a discrete domain with 8, 16, 32 and 64 points respectively, using ${\rm CFL}=1/2$, during one crossing time. For the spectral decomposition of the unknown function, we use $M=3$,  which is equivalent to consider the three lowest frequency modes in the truncated Fourier Series expansion of Eq. (\ref{eq:spectral}). We present the solutions: i) constructed with the traditional finite differences method of Eq. (\ref{eq:TridiagonalAdvection}), ii) using the SVF using the Cost Function (\ref{eq:CFadvection}) and iii) using the SEF. 

For the convergence test in Fig. \ref{fig:ConvergenceAdvection} we use the method described in the Appendix to verify that the solutions converge to the solution in the continuum. Since the discretization operator (\ref{eq:CNgenericOriginal}) and the Crank-Nicolson average are second order accurate, then ideally second order convergence of the numerical solution to the one in the continuum is expected. Basically we verify that numerical solutions calculated in numerical domains with different $N$, specifically increasing $N$ by factors of two, or equivalently decreasing $\Delta x$ by factors of two, approach quadratically to the solution in the continuum domain. 
Formally, in order to show convergence, we calculate the $L_1$ norm of the error of the numerical solution $u^{({\rm n})}{}^{n}_{i}$ at time $n$, calculated with ${\rm n}=3,4,5,6$ qbits, measured with respect to the exact solution $u^{({\rm n})}_{\rm ex}{}^{n}_{i}$. This norm is approximated with a numerical integration on the numerical domain $D_d$ given by 
$L_1(e^{{\rm (n)}n}) = \sum^{N_{\rm n}-1}_{i=0}~ |e^{({\rm n})}{}^{n}_{i}| ~\Delta x_{\rm n}$, where 
$e^{({\rm n})}{}^{n}_{i} = u^{({\rm n})}{}^{n}_{i} - u^{({\rm n})}_{\rm ex}{}^{n}_{i}$ is the error of the numerical solution. For ${\rm n}=4,5,6$, with the traditional and SVF methods, the factor of the norm between consecutive ${\rm n}$ is approximately 4, which indicates second order convergence, whereas for ${\rm n}=3$ the norm is not approximately four times bigger as with ${\rm n}=4$, indicating that in such case the solution is not within the convergence regime. We then say that the numerical solutions with these two methods is convergent for ${\rm n}=4,5,6$.

Concerning the solution using the SVF method, the solution is not convergent, and in fact is not numerically consistent, that is, for higher resolution the error sometimes is bigger than with coarse resolution. The solution in fact depends on the number of experiments and gives different results for different runs because it depends on the randomness of the sampling, and in the Figure we show a case in which the solution performs reasonably well. In summary, the classical method is well known to converge with second order, since discretization (\ref{eq:TridiagonalAdvection}) is second order accurate. Observe that the SVF also converges with second order. Finally, the SEF converges only in a very small time-window at the beginning, and later on it has problems even of numerical consistency.

\section{Application of the MoL:  The Wave Equation}
\label{subsec:wequation}

The initial value problem to be solved is defined on $D = [0,1] \times [0,1]$ by the wave equation

\begin{equation}
\partial^2_t \phi - \partial^2_x \phi=0,\nonumber 
\end{equation}

\noindent provided initial conditions $\phi_0(x)=\phi(x,0)$ and $\partial_t \phi(x,0)$. Instead of solving directly the second order equation, we define an equivalent problem that uses first order variables $P:=\partial_t \phi$ and $Q:=\partial_x \phi$, that evolve according to the following set of first order coupled equations:

\begin{eqnarray}
\partial_t P &=& \partial_x Q,\label{eq:we1storderP}\\
\partial_t Q &=& \partial_x P,\label{eq:we1storder}
\end{eqnarray}

\noindent with initial conditions $P_0(x)=P(x,0)$ and $Q_0(x)=Q(x,0)$, useful to illustrate how to solve a system of coupled equations with the MoL. We solve this formulation of the wave equation for an initially time-symmetric Gaussian pulse, that is $\phi_0(x)=\exp(-(x-x_0)^2/\sigma^2)$, with $\partial_t \phi(x,0) = 0$ with $x_0 = 0.5$ and $\sigma = 0.15$. These initial conditions translate to the first order variables as 

\begin{eqnarray}
P_0(x)&=& 0 ,\nonumber\\
Q_0(x)&=&-\frac{2(x-x_0)}{\sigma^2}e^{-\frac{(x-x_0)^2}{\sigma^2}}.\nonumber 
\end{eqnarray}

\noindent Finally, the solution is not obtained until the original unknown function $\phi$ is integrated, which we do from the definition of $P=\partial_t \phi$. Notice however that this can be written as an evolution equation for $\phi$, sourced by the value of $P$ at all times. This is the reason why we consider the complete system to be solved as the two equations (\ref{eq:we1storderP})-(\ref{eq:we1storder}) simultaneously with

\begin{equation}
\partial_t \phi= P,\label{eq:we1storderphi}
\end{equation}

\noindent which allows the integration of $\phi$ as soon as the value of $P$ is known at all times and intermediate steps during the evolution. We solve this three equations problem using the Method of Lines as described above, using the RK2 in Eqs. (\ref{eq:RK2a})-(\ref{eq:RK2b}), which can be generalized to solve a system of any number of first order equations. In particular, for the equations (\ref{eq:we1storderP}),(\ref{eq:we1storder}),(\ref{eq:we1storderphi}) for $P$, $Q$ and $\phi$, the two RK2 steps are given by

\begin{eqnarray}
\left(
\begin{array}{c}
P^*_i\\
\\
Q^*_i\\
\\
\phi^*_i \\
\end{array} \right) &=& \left(
\begin{array}{c}
P^n_i\\
\\
Q^n_i\\
\\
\phi^n_i
\end{array} \right) + \frac{\Delta t}{2\Delta x} \left(
\begin{array}{c}
Q^n_{i+1} - Q^n_{i-1}\\
\\
P^n_{i+1} - P^n_{i-1}\\
\\
2\Delta x P^n_i \\
\end{array} \right) \nonumber\\
\left(
\begin{array}{c}
P^{n+1}_i\\
\\
Q^{n+1}_i\\
\\
\phi^{n+1}_i\\
\end{array} \right) &=& \frac{1}{2}\left(
\begin{array}{c}
P^n_i + P^*_i\\
\\
Q^n_i + Q^*_i\\
\\
\phi^n_i + \phi^*_i\\
\end{array} \right) + \frac{\Delta t}{4\Delta x} \left(
\begin{array}{c}
Q^*_{i+1} - Q^*_{i-1}\\
\\
P^*_{i+1} - P^*_{i-1}\\
\\
2\Delta x P^*_i \\
\end{array} \right) .\nonumber
\end{eqnarray}

\noindent For the implementation of the VQA one has to minimize six costs functions. For the first Euler step, there are three independent cost functions derived from Eq. (\ref{eq:CFgeneralRK2a}), one for each of the three functions $P$, $Q$ and $\phi$:
\begin{eqnarray}
    CF_{\text{RK2-}a,P} &=& \bracketPaco{P^*}{P^*} - 2 {\rm Re} \{ \bracketPaco{P^*}{\tilde{P}} \} \nonumber \\
    &-& \frac{\Delta t}{\Delta x} {\rm Re} \{ \langle P^*|\circledplus - \circledminus | \tilde{Q}\rangle\}\nonumber \\
    CF_{\text{RK2-}a,Q} &=& \bracketPaco{Q^*}{Q^*} - 2 {\rm Re} \{ \bracketPaco{Q^*}{\tilde{Q}} \} \nonumber \\
    &-& \frac{\Delta t}{\Delta x} {\rm Re} \{ \langle Q^*|\circledplus - \circledminus | \tilde{P}\rangle\} \nonumber\\
    CF_{\text{RK2-}a,\phi} &=& \bracketPaco{\phi^*}{\phi^*} - 2 Re \{ \bracketPaco{\phi^*}{\tilde{\phi}} \} \nonumber\\
    &-& 2\Delta t {\rm Re} \{ \bracketPaco{\phi^*}{\tilde{P}}\}.\nonumber
\end{eqnarray}

\noindent These cost functions are minimized to find the $\ket{P^*}$, $\ket{Q^*}$ and $\ket{\phi^*}$ auxiliary states. For the second RK2 step (\ref{eq:CFgeneralRK2b}), we find the following cost functions:


\begin{figure}
\centering
\includegraphics[width=8cm]{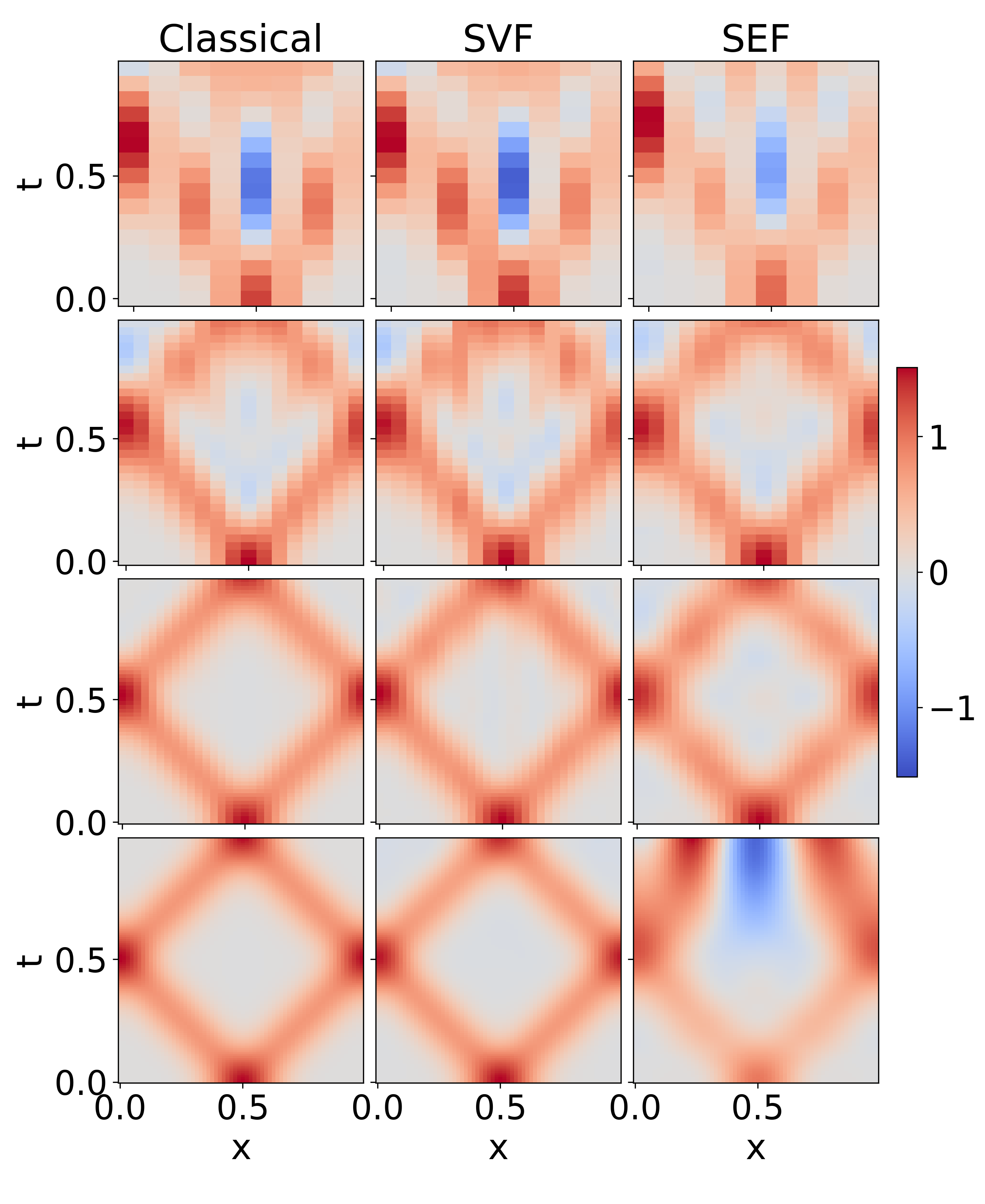}
\includegraphics[width=7.5cm]{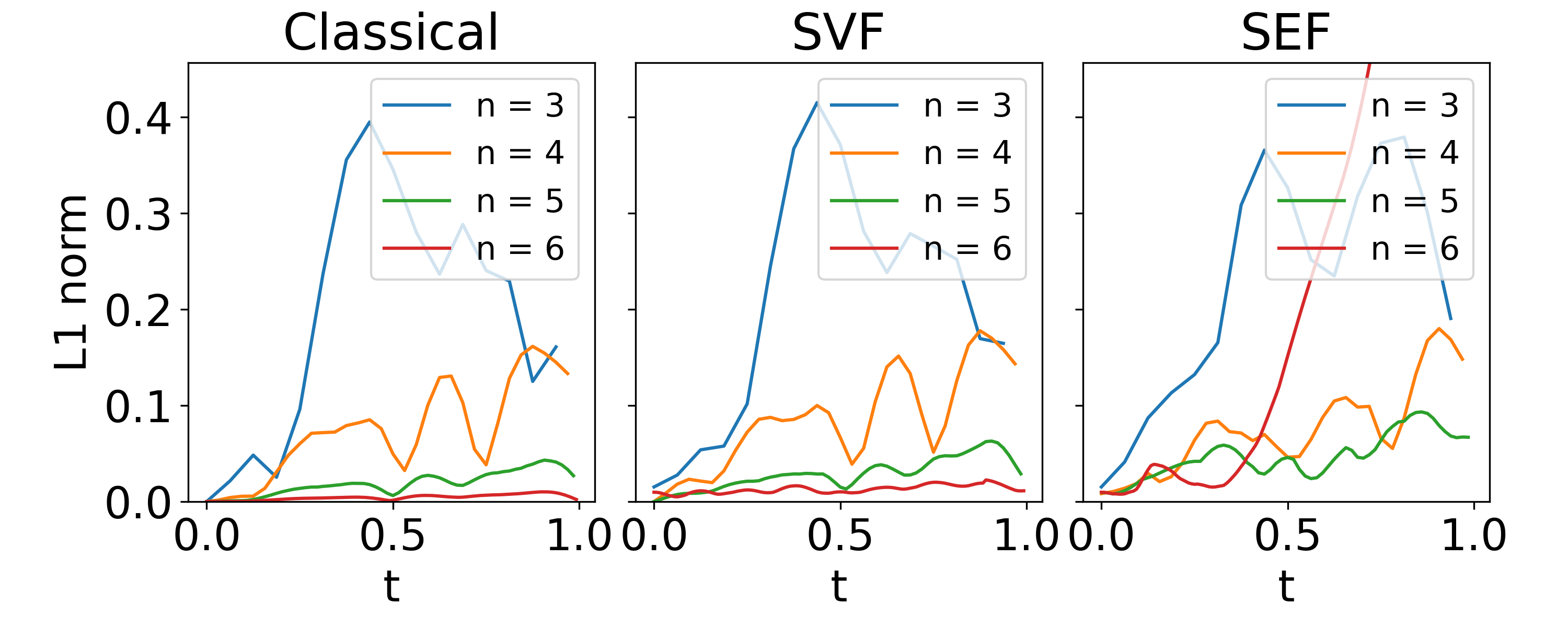}
\caption{From left to right we show the solution of the wave equation using the classical method, the SVF and SEF  respectively. At the first, second, third and fourth rows we show the solution calculated with 3, 4, 5 and 6 qubits respectively, using $M=7$ modes for ${\rm n} > 3$ and $M=3$ for ${\rm n}=3$. 
 The colors indicate the amplitude of the wave function $\phi_i^n$, an initially centered Gaussian of amplitude one that splits into two pulses moving outwards that reenter through the periodic domain. 
At the bottom we show a convergence test of the $L_1$ norm of the error as function of time using the four resolutions. This test indicates that the traditional method and the SVF approach converge with second order, whereas the SEF does not converge, instead it is very sensitive and strongly depends on the number of experiments; in this test we use $T = 10^{9}$ experiments and still convergence and consistency are lost soon after initial time.}
\label{fig:ConvergenceWave}
\end{figure}

\begin{eqnarray}
    CF_{\text{RK2-}b,P} &=& \bracketPaco{P}{P} -  {\rm Re} \{ \bracketPaco{P}{\tilde{P}} + \bracketPaco{P}{P^*}\} \nonumber \\
    &-& \frac{\Delta t}{2\Delta x} {\rm Re} \{ \langle P|\circledplus - \circledminus | Q^*\rangle\} \nonumber\\
    CF_{\text{RK2-}b,Q} &=& \bracketPaco{Q}{Q} -  {\rm Re} \{ \bracketPaco{Q}{\tilde{Q}} + \bracketPaco{Q}{Q^*}\} \nonumber \\
    &-& \frac{\Delta t}{2\Delta x} {\rm Re} \{ \langle Q|\circledplus - \circledminus | P^*\rangle\} \nonumber\\
    CF_{\text{RK2-}b,\phi} &=& \bracketPaco{\phi}{\phi} -  {\rm Re} \{ \bracketPaco{\phi}{\tilde{\phi}} + \bracketPaco{\phi}{\phi^*}\} \nonumber \\
    &-& \Delta t Re \{ \bracketPaco{\phi}{P^*}\}.\nonumber
\end{eqnarray}

\noindent These cost functions are minimized in the following order. First minimize $CF_{\text{RK2-}a,P}$ to find the parameters that define $\ket{P^*}$. Second, minimize $CF_{\text{RK2-}a,Q} $ in order to find independently the parameters that best define $\ket{Q^*}$. And third, minimize $CF_{\text{RK2-}a,\phi}$ to find $\ket{\phi^*}$. Once $\ket{P^*}$, $\ket{Q^*}$ and $\ket{\phi^*}$ are determined, we minimize $CF_{\text{RK2-}b,P}$, $CF_{\text{RK2-}b,Q}$ and $CF_{\text{RK2-}b,\phi}$ independently to find the parameters that determine $|P(t^{n+1})\rangle$, $|Q(t^{n+1})\rangle$ and $|\phi(t^{n+1})\rangle$.

The evolution obtained using various resolutions during one crossing time are shown in Figure \ref{fig:ConvergenceWave}. First notice that the initial Gaussian pulse splits into two half-height Gaussians, one moving to the left and another one moving to the right as expected from the initial time symmetry, until they superpose again at $t=1$. The SVF and SEF  meet very well the properties of the traditional MoL using 3, 4 and 5 qubits, whereas for 6 qubits the solution with the SEF method deteriorates. Likewise for the Advection equation, the convergence tests in the Figure show that the convergence regime for the Classical MoL starts from 4 qubits on, as well as when using the SEF method, however the solution with the SVF method does not converge when increasing the number of qubits, and is not numerically consistent; notice also that the behavior of the SEF solution is different for different runs because the cost function has stochastic behavior and gives unpredictably different results each time it is calculated.

\section{Application:  Burgers Equation}
\label{subsec:bequation}

We now investigate how the method behaves for an equation that tends to develop discontinuities \cite{leveque1992numerical}. Unlike some examples that use non-linear equations to illustrate some evolution methods \cite{BlackScholes,princeton,Mocz_2021,AppRusosNonLinear} but do not develop discontinuities, Burgers equation does, and certainly adds some limits to the applicability of the spectral expansion of functions. Knowing in advance that the solution to this equation forms discontinuities, we anticipate and formulate the IVP on $D$ using the Burgers equation with viscosity $\nu$ as follows:

\begin{equation}
\partial_t u + u\partial_x u = \nu \partial^2_x u,\label{eq:BurgersNU}
\end{equation}

\noindent for some initial conditions $u(x,0)= u_0(x)$. A first order algorithm to solve this problem is presented in \cite{Lubasch_2020}, although no solutions were computed. Here we implement a second order version of such algorithm using the MoL. The semi-discrete version of the equation using second order stencils becomes:
\begin{equation}
    \partial_t u^n_i = -u^n_i \left( \frac{u^n_{i+1} - u^n_{i-1}}{2\Delta x}\right) + \nu \left( \frac{u^n_{i+1} - u^n_i + u^n_{i-1}}{\Delta x^2}\right),\nonumber
\end{equation}

\noindent which, translated into the quantum algorithm language reads:

\begin{equation}
    \partial_t \ket{u} = -\frac{1}{2\Delta x}\hat{u} ( \circledplus - \circledminus)\ket{u} + \frac{\nu}{\Delta x^2} ( \circledplus - 2\mathbf{I} + \circledminus )\ket{u},\nonumber
\end{equation}

\noindent where $\hat{u}$ is a new linear operator, which is Hermitian instead of unitary. This operator is defined by the diagonal matrix ${\rm Diag}(u^n_{0},u^n_{1},\dots,u^n_{N-1})$. With the given right hand side, the cost functions used to do the first and second steps of the RK2 method (\ref{eq:RK2a})-(\ref{eq:RK2b}) are respectively:

\begin{figure}
\includegraphics[width=8.5cm]{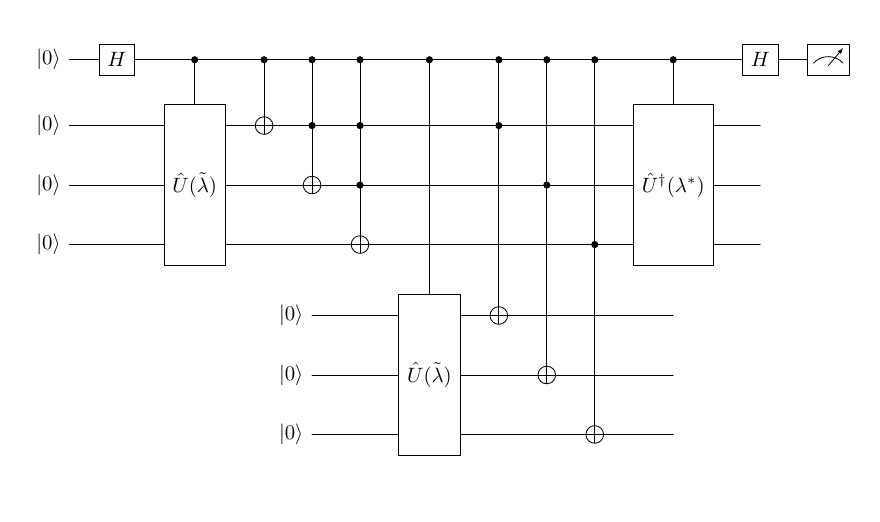}
\caption{Circuit that evaluates the product 
$ {\rm Re} \{ \langle u^*| \hat{\tilde{u}} \oplus |\tilde{u} \rangle \} = {\rm Re} \{ \langle 0| \hat{U}^{\dagger}(\lambda^*)\hat{\tilde{u}} \oplus \hat{U}(\tilde{\lambda}) |0\rangle\}$ in Eq. (\ref{eq:CFburgersA}). To find the value of the product, the circuit must be executed multiple times and the ancilla qubit has to be measured. The desired result will be the probability of measuring $0$ minus the probability of measuring $1$ if all the functions are normalized. In this example ${\rm n} = 3$.}
\label{circuit:QNPU}
\end{figure}

\begin{figure}
\centering
\includegraphics[width=8cm]{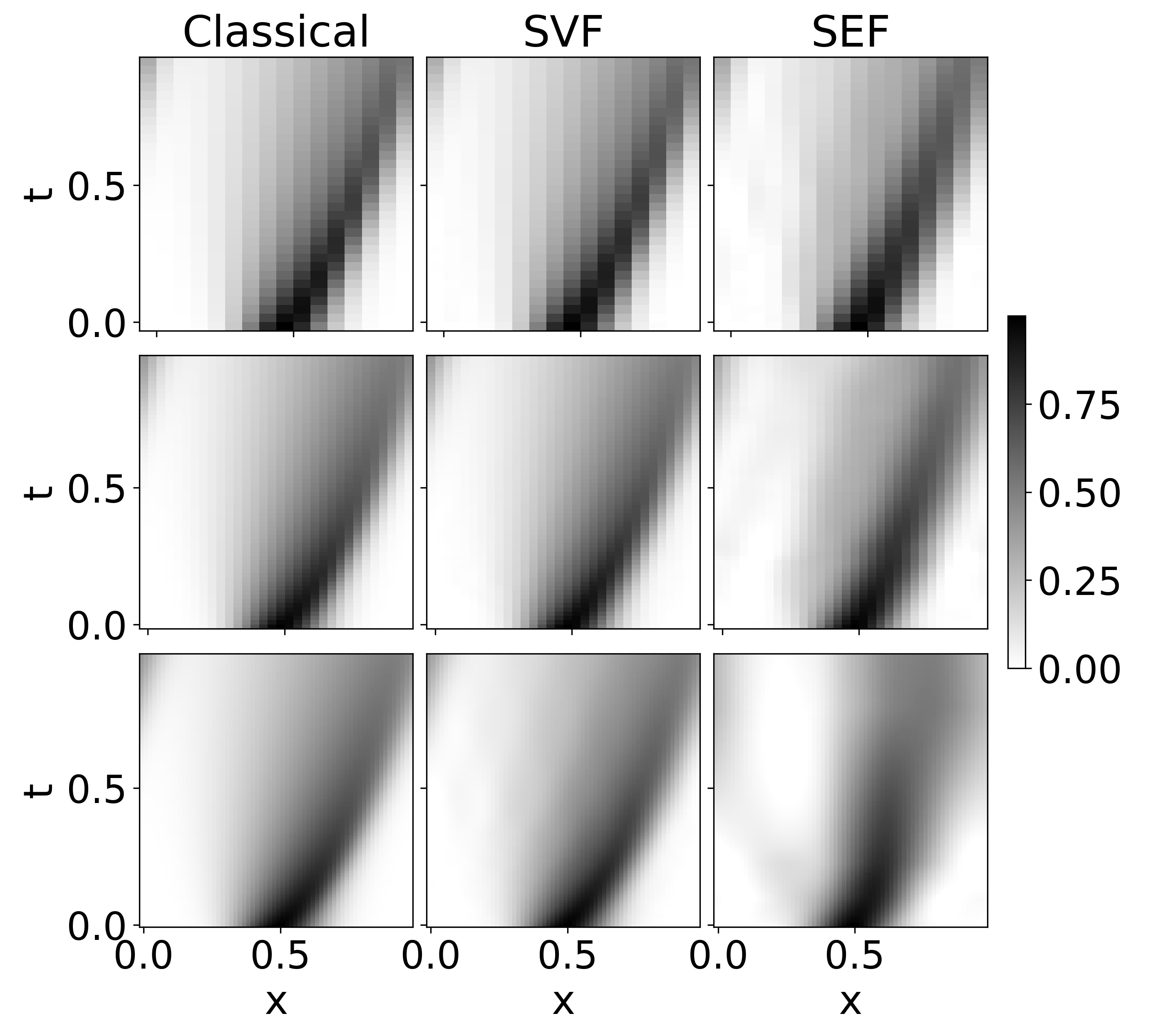}
\includegraphics[width=7.5cm]{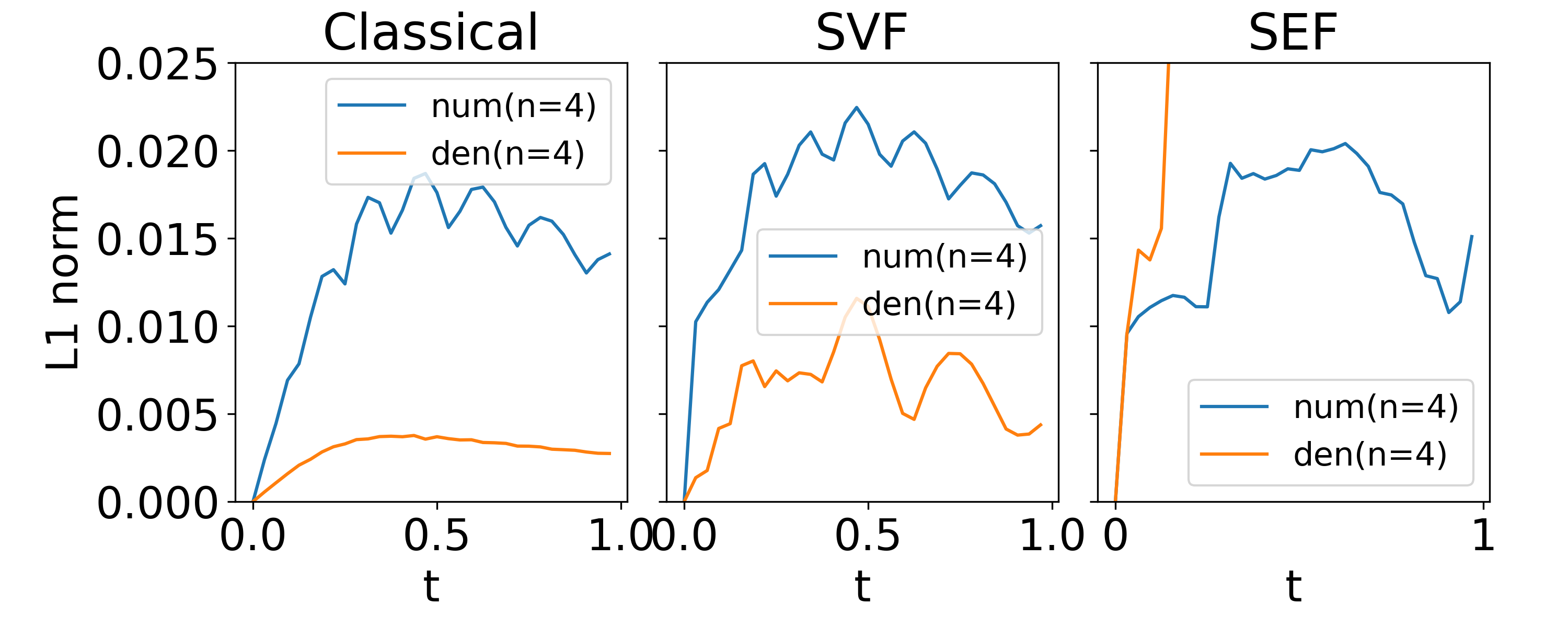}
\caption{From left to right we show the solution of Burgers equation (\ref{eq:BurgersNU}) for an initial Gaussian profile, using the classical, the SVF and SEF formalisms respectively. At the first, second and third rows we show the solution calculated with 4, 5 and 6 qubits respectively, using $M=7$ frequencies in the ansatz and $T=10^{12}$ shots.
 The grey scale indicate the amplitude of the numerical solution $u_i^n$, an initially centered Gaussian of amplitude one that moves towards the right side and tends to form a discontinuity. In the bottom of the Figure, a self-convergence test is shown, according to the description in the Appendix. It can be seen that the classical solution self-converges with second order, the SVF self-converges with a convergence factor between 1 and 2, the SEF doesn't self-converge.}
\label{fig:Burgers}
\end{figure}

\begin{eqnarray}
    CF_{\text{RK2-}a} &=& \bracketPaco{u^*}{u^*} - 2 {\rm Re} \{ \bracketPaco{u^*}{\tilde{u}} \} \label{eq:CFburgersA}\\ 
    &+& \frac{\Delta t}{\Delta x} {\rm Re} \{ \sandwichPaco{u^*}{\hat{\tilde{u}}(\circledplus - \circledminus)}{\tilde{{u}}}\}, \nonumber\\
    &-& \nu\frac{2\Delta t}{\Delta x^2} {\rm Re} \{ \sandwichPaco{u^*}{\circledplus -2\mathbf{I} + \circledminus}{\tilde{{u}}}\}, \nonumber\\
    CF_{\text{RK2-}b} &=& \bracketPaco{u}{u} -  {\rm Re} \{ \bracketPaco{u}{\tilde{u}} + \bracketPaco{u}{u^*}\} \nonumber\\
    &+& \frac{\Delta t}{2\Delta x} {\rm Re} \{ \sandwichPaco{u}{\hat{u^*}(\circledplus - \circledminus)}{u^*}\}, \nonumber\\
    &-& \nu\frac{\Delta t}{\Delta x^2} {\rm Re} \{ \sandwichPaco{u}{\circledplus -2\mathbf{I} + \circledminus}{u^*}\}. \nonumber
\end{eqnarray}

\noindent This cost function includes new products like ${\rm Re} \{ \sandwichPaco{u}{\hat{u^*}\circledplus}{u^*} \}$ that cannot be computed using the quantum circuits used before. To address this issue, the Quantum Nonlinear Processing Unit described in \cite{Lubasch_2020} provides the recipe to deal with the type of non-linearities of Burgers equation. The appropriate circuit is shown in Figure \ref{circuit:QNPU}. The three gates after the first $\hat{U}(\tilde{\lambda})$ gate are the gate representation of the $\circledplus$ operator. If their order is reversed, then the circuit can compute the product ${\rm Re} \{ \sandwichPaco{u}{\hat{u^*}\circledminus}{u^*} \}$. By switching the parameters of the $\hat{U}$ gates, multiple products can also be computed. 

We solve this IVP in the domain $D=[0,1]\times [0,1]$, for initial conditions $u_0(x)=e^{-(x-x_0)^2/\sigma^2}$, with $x_0=0.5$ and $\sigma=0.15$. We also set $\nu = 0.0125$ to keep a stable evolution. The solution can be seen in Figure \ref{fig:Burgers}. From top to bottom we use 4, 5 and 6 qubits.

Various observations are to be commented. First is that the solution using the SVF performs well. The Fourier ansatz is able to represent the sharp peak of the solution with only $M = 7$ frequencies. Concerning the SEF it happens the same as in the other cases, namely, the solution degrades with the number of iterations. Notice also that, since the exact solution is not known we cannot present a convergence test as done in the previous two examples. Instead a self-convergence test is in turn, which consists of verifying whether when increasing resolution by factors of two, the numerical solution converges to itself in the continuum limit as described in the Appendix. Second order self-convergence would be the ideal convergence rate, since the discretization of derivative  operators and time-integration from Section II are second order accurate. Formally, in the bottom of Figure \ref{fig:Burgers} we show in blue the difference between the numerical solutions with ${\rm n}=4$ and ${\rm n}=5$,  which would be the numerator of Eq. (\ref{eq:selfconvergcenfactorPDE}) for ${\rm n}=4$, while in yellow we draw the difference between the numerical solutions with ${\rm n}=5$ and ${\rm n}=6$, that is,  the denominator of Eq. (\ref{eq:selfconvergcenfactorPDE}) for ${\rm n}=4$. Since the factor between the two curves is close to $2^2$ for the classical method we say that the numerical solution self-converges with order two, whereas the SVF self-converges with an order between one and two since the factor lies between $2^1$ and $2^2$; finally the SEF does not self-converge.


\section{Complexity Analysis}
\label{sec:analysis}

In this section, we briefly analyse the computational complexity of the algorithm. We start by calculating the complexity of computing the cost function once. The cost functions employed in this work have a constant number of terms, which implies a constant number of circuits, which does not depend on n. Each circuit consists of the ansatz, used once or multiple times, along with intermediary gates. The ansatz requires ${\cal O}({\rm n}^2 + M)$ gates, while the intermediary operators ($\circledplus$, $\circledminus$, etc) require at most ${\cal O}({\rm n}^2)$ gates. Therefore, the total complexity of the circuits is ${\cal O}({\rm n}^2 + M)$.

Each circuit is executed $T$ times, depending on the desired precision. This quantity is independent of the number of qubits. Therefore, the complexity of evaluating the cost function, considering that the number of terms remains constant, is ${\cal O}(T[{\rm n}^2 + M])$.

\begin{figure}
\centering
\includegraphics[width=7.0cm]{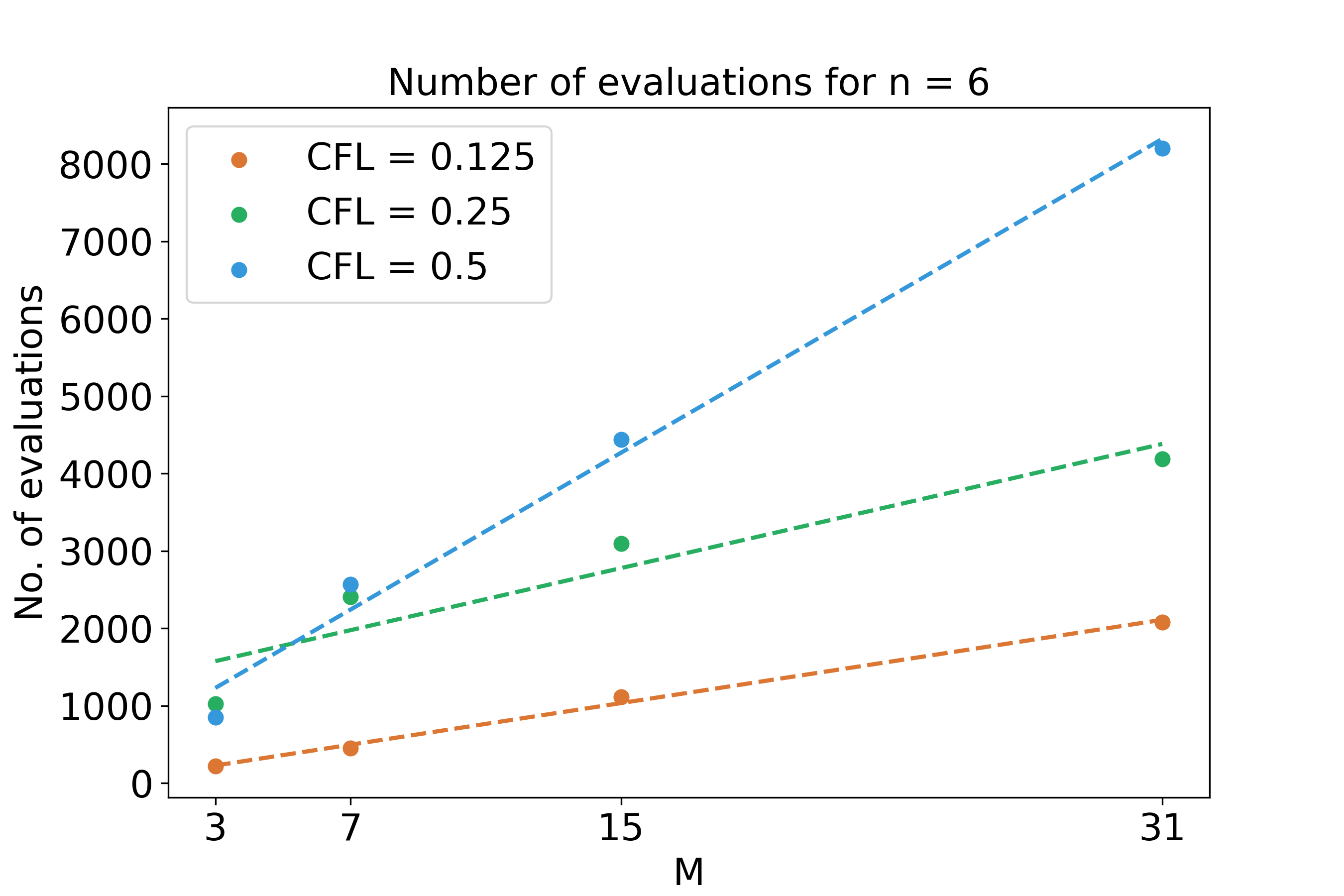}
\caption{Average number of cost function calls per time step during the evolution of the Advection equation. The average is taken over a crossing time: from $t=0$ to $t=1$. In this case ${\rm n} = 6$ and $\rm{m}$ takes over the values: $2$, $3$, $4$ and $5$. Three values for the Courant Friedrichs Lewy (CFL) factor are used. We can see that the number of evaluations grows linearly with the number of terms in the Fourier series: $M$. It can also be seen that it is also proportional to the CFL.}
\label{fig:EvaluationsA}
\end{figure}

After evaluating the cost function, the {\tt Nelder-Mead} method will perform an adjustment of parameters and requires ${\cal O}(M)$ operations, which does not affect the complexity because it is done in series with the computation of the cost function. Finally, in Figure \ref{fig:EvaluationsA} it can be seen that the number of evaluations required to perform one time step is proportional to the number of frequency modes $M$. Therefore, the time complexity of the algorithm required to perform one time step is:
\begin{equation}
    {\cal O}(TM[{\rm n}^2 + M]),
    \label{eq:complexity}
\end{equation}
which is an exponential speed-up over the classical algorithm which has complexity of order ${\cal O}(2^{\rm n})$.

One interesting fact about our implementation is that, as it can be seen in Figure \ref{fig:EvaluationsA}, the number of evaluations of the cost function is proportional to the ${\rm CFL}$ factor for the Advection equation, which is our simplest example. This means that, increasing the time resolution does not increase the number of operations, which is a difference with respect to the classical numerical algorithm.

\section{Final comments}
\label{sec:conclusions}

We used the VQA to solve initial value problems in 1+1 dimensions on a periodic domain. We assumed the unknowns were described with a truncated Fourier basis with a small number of terms. The evolution was carried out using two basic numerical evolution scehmes: the implicit Crank-Nicolson exemplified with the Advection Equation, and the Method of Lines applied to solve the Wave Equation written as a system of first order coupled equations and the viscous Burgers equation. We use these examples to investigate the capability and performance of the VQA with the two evolution schemes. 

Our case of control to compare with was the classical solution. It is the guide to determine how well the solutions with SVF and SEF methods do. We notice that convergence and self-convergence is as expected, but the minimum number of points to lie within the convergence regime is 16. This is equivalent to use at least 4 qubits to enter the convergence regime. A numerical solution is correct or at least reliable if convergence is achieved, otherwise numerical solutions are worthless. Fortunately, this convergence regime holds for SVF and SEF, at least for a finite time interval.

The convergence of the SVF in the Advection and wave equations shows that the Fourier expansion ansatz used is sufficient to achieve convergence in equations that do not promote the formation of discontinuities. On the other hand, in the Burgers equation case, the characteristic lines tend to converge, and despite the viscosity, the unknown function acquires a sharp profile. However, self-convergence was still achieved between first and second order. It is important to take into account that $4$ qubits or more must be used in order to obtain convergence.

Concerning the SEF, even though the equations solved in this paper are the simplest of their kind, our convergence tests indicate that the VQA method under-performs when sampling errors are taken into account. Solutions are only consistent during a small time window: $t \in [0,0.1]$; the rest of the evolution highly depends on the number of shots used to execute the quantum circuits. For high resolutions, since $\Delta t$ shrinks, the cost function has to be minimized more times and the sampling error at late times accumulates; this causes the unknown functions to get distorted. Also notice that in order to obtain consistent numerical solutions within a crossing time, the number of shots used for the quantum circuits is of order of $10^{11}$, which is high compared to the $20,000$ shots available in today's quantum computers \cite{qiskit}. After all, as far as we can tell, there are no comparable convergence tests for similar extra simple implementations that could validate the use of VQAs to solve IVPs, specially when sampling errors are taken into account. This of course questions the possibility of solving IVPs on real quantum hardware, where besides the sampling, decoherence errors must be handled. To address this issue, other optimizers, better in the sense that they find the global minimum of stochastic functions with a reduced number of experiments, must be tested for example those presented in \cite{StocacticOp}.

Byproducts of our analysis are: 1) the construction of cost functions useful for the Crank Nicholson method (\ref{eq:CFgeneralCN}) and for the Method of Lines for a system of coupled equations (\ref{eq:CFgeneralRK2a})-(\ref{eq:CFgeneralRK2b}), that may be useful as they are, or generalized for more stable and higher order time integrators; 2) a convergent solution of a system of three coupled PDEs was obtained; 3) the fast SVF and SEF were developed, although these only work when using the truncated Fourier series ansatz, they can be used to test optimizers rapidly. 

The approach in this paper is expected to be expanded towards the solution of problems in non-periodic domains, that could include the development of non-periodic boundary conditions as advanced in \cite{NonPeriodicBCs,BCs2024}.

\section*{Acknowledgments}
This work is supported by CIC-UMSNH Grant  No. 4.9, CONAHCyT Ciencias de Frontera Grant No. Sinergias/304001 and 
Laboratorio Nacional de C\'omputo de Alto Desempe\~no Grant No. 1-2024.


\bibliography{QC2}

\appendix
\section{Convergence and Self-Convergence of numerical solutions}
\label{app:convergence}

The formulation of the IVP on a periodic domain defines an equal number $N$ of points and cells used to discretize the spatial coordinate. The fact that adding one qubit doubles the resolution is helpful to define  convergence tests. For an IVP one has to check convergence of numerical solutions at all points of the numerical space-time domain, which means that at all times $t^n$ and at all points $x_i$ of $D_d$, the solution should converge to the exact solution in the continuum limit.

{\it Convergence test.} This test needs the numerical solution of the IVP calculated in domains that use different resolutions, for which, following \cite{fsguzman} we define a hierarchy of numerical domains $D^{{\rm n}}_{d}$ constructed with different spatial resolutions where numerical solutions are calculated. The number of qubits ${\rm n} = 3,4,5,6$ defines the spatial resolution of $D_d$,  thus we denote this discrete domain for each n as $D^{{\rm n}}_{d}$ with the following properties:

\begin{equation}
\begin{tabular}{c|c|c}
n & No. of cells $N$ & Spatial Resolution\\\hline
3   &   $N_3=2^3$ & $\Delta x_3=1/2^3$\\
4   &   $N_4=2^4$ & $\Delta x_4=1/2^4$\\
5   &   $N_5=2^5$ & $\Delta x_5=1/2^5$\\
6   &   $N_6=2^6$ & $\Delta x_6=1/2^6$
\end{tabular}\label{tab:tabla}
\end{equation}

\noindent  We calculate the numerical solutions $u^{({\rm n})}{}^{n}_{i}$, ${\rm n}=3,4,5,6$ in the four domains $D^{{\rm n}}_{d}$, that use resolution $\Delta x_{\rm n}$ and $N_{\rm n}$ cells. Notice that each numerical domain is a subset of the next finer one, that is $D^{3}_{d} \subset D^{4}_{d} \subset D^{5}_{d} \subset D^{6}_{d}$, which in turn is a subset of $D$. 

Convergence tests also need the exact solution of the problem, that we denote for each resolution by $u^{({\rm n})}_{\rm ex}{}^{n}_{i}$ at the point $(x_i,t^n)\in D^{{\rm n}}_{d}$ . Then the difference between the numerical and exact solutions is the error of the solution at each point defined as

\begin{equation}
e^{({\rm n})}{}^{n}_{i} = u^{({\rm n})}{}^{n}_{i} - u^{({\rm n})}_{\rm ex}{}^{n}_{i}. \label{eq:errorpde}
\end{equation}

\noindent When this error decreases with increasing resolution, the solution is {\it consistent}; when it decreases with a rate according with the error of the method, the solution is {\it convergent} \cite{thomasNPDE}.

Assume now that the discretization of the evolution equation we solve is $k-$th order accurate. Then take any pair of numerical solutions with consecutive resolutions, then they can be expressed in terms of the exact solution with the addition of an error term:

\begin{eqnarray}
u^{({\rm n})}{}^{n}_{i} &=& u^{({\rm n})}_{\rm ex}{}^{n}_{i}
 + E(x_i) (\Delta x {}_{{\rm n}})^k,\nonumber\\
u^{({\rm n}+1)}{}^{n}_{i} &=& u^{({\rm n+1})}_{\rm ex}{}^{n}_{i} + E(x_i) \left( \frac{\Delta x{}_{{\rm n}}}{2} \right)^k,\label{eq:convergencePDE}
\end{eqnarray}

\noindent at point $(x_i,t^n)\in  D^{{\rm n}}_{d} \subset D^{{\rm n+1}}_{d} $, and the error term uses the fact that consecutive resolutions are being doubled as seen in the domain's properties list (\ref{tab:tabla}).  It is said that the numerical solution converges to the exact solution at point $(x_i,t^n)$ if the following relation holds:

\begin{equation}
C:=\frac{e^{({\rm n})}{}^{n}_{i}}{e^{({\rm n}+1)}{}^{n}_{i}}=
\frac{u^{({\rm n})}{}^{n}_{i} - u^{({\rm n})}_{\rm ex}{}^{n}_{i}}{ u^{({\rm n}+1)}{}^{n}_{i} - u^{({\rm n+1})}_{\rm ex}{}^{n}_{i} } 
\simeq  \frac{\Delta x{}_{{\rm n}}^k}{\left(\frac{\Delta x{}_{\rm n}}{2}\right)^k} = 2^k,
\label{eq:convergcenfactorPDE}
\end{equation}

\noindent at least approximately, where $C$ is the {\it convergence factor}. It is said that the numerical solutions converge with order $k$, exactly the accuracy of the discretization errors. In general the solutions may converge with order $l$ if $C\simeq 2^l$.

For the particular case of our solutions, in all cases time and spatial discretization, for both evolution schemes Crank-Nicolson and MoL, are second order accurate, thus $k=2$ and the convergence factor for a convergent solution should be  $C \simeq 2^2$.

Now, this convergence factor should be calculated at each point of the numerical domain, which is not practical. What we do is use Lax Theorem \cite{thomasNPDE} and calculate a norm, the $L_1$ in our case (it could have been $L_2$ or $L_{\infty}$), of the error (\ref{eq:errorpde}) at time $t^n$ as follows:

\begin{eqnarray}
L_1(e^{{\rm (n)}n}) &= \sum^{N_{\rm n}-1}_{i=0}~ |e^{({\rm n})}{}^{n}_{i}| ~\Delta x_{\rm n},\label{eq:L1normintegral}
\end{eqnarray}

\noindent that uses the trapezoidal rule on the periodic domain. Lax Theorem guarantees that convergence of the norm implies convergence at each spatial point, that is $L_1(e^{{\rm (n)}n})/L_1(e^{{\rm (n+1)}n})$ should be $\simeq 2^2$ if the numerical solution converges. In order to monitor the convergence of this norm, we calculate $L_1(e^{({\rm n})n})$ during the evolution for all the resolutions and monitor the factor between them. 

At the bottom row of Figures \ref{fig:ConvergenceAdvection} and \ref{fig:ConvergenceWave} we show the $L_1$ norm of the error of numerical solutions using ${\rm n}=3,4,5,6$ with blue, yellow, green and red lines respectively. We say that solutions converge when the factor between consecutive lines is approximately $2^2$. 

A word on {\it consistency} is in turn. Notice that in these Figures the results obtained with the SEF method, from a finite time on, the errors with finer resolution are bigger than those with coarser resolution; we say that numerical solutions in that case are not only non-convergent, but also non-consistent.

{\it Self-Convergence test.} When the exact solution is unknown, one can define self-convergence using the numerical solution calculated with a third resolution:

\begin{equation}
u^{({\rm n}+2)}{}^{n}_{i} = 
u^{({\rm n+2})}_{\rm ex}{}^{n}_{i}+ E(x_i) \left( \frac{\Delta x{}_{\rm n}}{4} \right)^k ,
\end{equation}

\noindent that together with those in (\ref{eq:convergencePDE}) allow the following analysis. It is said that the numerical solution self-converges at point $(x_i,t^n)\in D^{{\rm n}}_{d} \subset D^{{\rm n+1}}_{d} \subset D^{{\rm n+2}}_{d}$, if the following relation between the three numerical solutions holds

\begin{eqnarray}
SC&:=&\frac{u^{({\rm n})}{}^{n}_{i} - u^{({\rm n}+1)}{}^{n}_{i}}{ u^{({\rm n}+1)}{}^{n}_{i} - u^{({\rm n}+2)}{}^{n}_{i} } \simeq  
\frac{(\Delta x{}_{{\rm n}})^k - \left( \frac{\Delta x{}_{{\rm n}}}{2}\right)^k}{\left( \frac{\Delta x{}_{{\rm n}}}{2}\right)^k - \left( \frac{\Delta x{}_{{\rm n}}}{4}\right)^k} \nonumber \\
&=& \frac{ 1 -\frac{1}{2^k} }{ \frac{1}{2^k} - \frac{1}{4^k}  } = 2^k
\label{eq:selfconvergcenfactorPDE}
\end{eqnarray}

\noindent where $SC$ is the {\it self-convergence factor}. Then it is said that numerical solutions self-converge with order $k$, which is the order of the discretization error. In general the solutions may self-converge with order $l$ if $SC\simeq 2^l$. In our case, since the discretization used in all the examples of this paper are second order, then $k=2$ and a numerical solution self-converges when $SC \simeq 4$.

Likewise in the convergence test, it is not practical to calculate this factor at each point. Then we calculate the $L_1$ norm of  the numerator $u^{({\rm n})}{}^{n}_{i} - u^{({\rm n}+1)}{}^{n}_{i}$ and denominator $u^{({\rm n}+1)}{}^{n}_{i} - u^{({\rm n}+2)}{}^{n}_{i}$ of (\ref{eq:selfconvergcenfactorPDE}) using the trapezoidal rule as in Eq. (\ref{eq:L1normintegral}). 
These scalars, the numerator and denominator are plotted in the bottom row of Figure \ref{fig:Burgers} as functions of time for ${\rm n} = 4$, the numerator in blue (num({\rm n} = 4)) and the denominator in yellow (den({\rm n} = 4)). In the classical solution, it can be seen that the ratio between the curves is approximately $4$; in the SVF, the ratio is between $2$ and $4$, which indicates the solutions do not self-converge with convergence of order two, but between order one and two; the SEF seems to anti-self-converge, because the difference between the solutions grows when resolution is increased.

\end{document}